\documentclass[11pt,a4paper]{article}

\usepackage[margin=2.5cm]{geometry}
\usepackage{graphicx}
\usepackage{amsmath,amssymb}
\usepackage{siunitx}
	\DeclareSIUnit{\angstrom}{\textup{\AA}}
\usepackage{booktabs}
\usepackage{caption}
\usepackage{subcaption}
\usepackage[colorlinks=true,allcolors=blue]{hyperref}
\usepackage{xcolor}
\usepackage{pdfpages}

\title{\textbf{Gallium phosphide on insulator for nanophotonics and quantum technologies}}

\author{
	Tobias Bucher$^{1,\#}$, Otto Arnold$^{1,\#}$, Muyi Yang$^{1,2,3}$, Zifei Zhang$^{1,2}$,\\
	Katsuya Tanaka$^{1,2}$, Annkathrin K\"ohler$^{1}$, Berit Marx-Glowna$^{4,5,6}$, Duk-Yong Choi$^{7}$,\\
	Isabelle Staude$^{1,2,3,8}$, Carsten Ronning$^{1,8}$
}

\date{}

\begin{document}
	
	\maketitle
	
	\begin{center}
		\small
		$^{1}$Institute of Solid State Physics, Friedrich Schiller University Jena, 07743 Jena, Germany\\
		$^{2}$Institute of Applied Physics, Friedrich Schiller University Jena, 07743 Jena, Germany\\[0.5em]
		$^{3}$Max Planck School of Photonics, Friedrich Schiller University Jena, 07745 Jena, Germany\\[0.5em]
		$^{4}$Helmholtz-Institut Jena, 07743 Jena, Germany\\[0.5em]
		$^{5}$GSI Helmholtzzentrum für Schwerionenforschung, 64291 Darmstadt, Germany\\[0.5em]
		$^{6}$Institute for Optics and Quantum Electronics, Friedrich Schiller University Jena, 07743 Jena, Germany\\[0.5em]
		$^{7}$Laser Physics Centre, Research School of Physics, Australian National University, Canberra, ACT, 2601, Australia\\[0.5em]
		$^{8}$Abbe Center of Photonics, Friedrich Schiller University Jena, 07745 Jena, Germany\\[0.5em]
		
		$^{\#}$Contributed equally to this study.\\[0.5em]
		$^{*}$Corresponding authors: tobias.bucher@uni-jena.de, otto.arnold@uni-jena.de
	\end{center}
	\begin{abstract}
		\noindent Gallium phosphide is a promising material platform for visible and near-infrared photonics and quantum technologies owing to its high refractive index, low optical absorption, and strong second-order nonlinearity. Here, we demonstrate the fabrication of GaP-on-insulator substrates by ion slicing. The splitting depth and exfoliation behavior of bulk GaP are tailored by controlling the He$^{+}$ ion implantation energy and fluence, enabling thin-film transfer onto amorphous substrates by anodic bonding and plasma-enhanced direct wafer bonding. Channeling Rutherford backscattering spectrometry and X-ray diffraction confirm that the transferred layers retain their single-crystalline structure, while implantation-induced disorder and optical absorption are substantially reduced by annealing at \qty{500}{\degreeCelsius} and subsequent polishing. The annealed films exhibit linear optical properties approaching those of bulk GaP. In addition, a (110)-oriented GaP thin film shows the characteristic polarization dependence expected from the zinc-blende second-order nonlinear susceptibility tensor, demonstrating a near-pristine second-order nonlinear response. This flexible fabrication approach enables the integration of high-quality single-crystalline GaP with variable orientation for free-space and integrated nanophotonics as well as nonlinear and quantum optical devices.
	\end{abstract}
	\section*{Introduction}\label{sec:intro}
	The miniaturization of quantum-optical systems into nanophotonic architectures is a key prerequisite for translating the second quantum revolution into scalable technologies \cite{Wang2020Integrated,Elshaari2020Hybrid}. Integrated platforms, including photonic integrated circuits and metasurfaces, provide the stability, reproducibility, and device density required to generate, manipulate, and detect quantum states of light within compact optical systems. Their performance and scalability, however, remain fundamentally constrained by the properties of the underlying materials \cite{Pelucchi2022Thepotential,Dutt2024Nonlinear}.\\
	Established platforms such as silicon, silicon nitride, and lithium niobate have enabled major advances in integrated photonics, yet each platform offers only part of the functionality required for fully integrated quantum-optical systems. Silicon and silicon nitride offer mature fabrication and low-loss optical confinement but, as centrosymmetric materials, lack an intrinsic bulk second-order nonlinearity \cite{Dutt2024Nonlinear}. Lithium niobate provides strong electro-optic functionality and a useful second-order nonlinearity, but its moderate refractive index limits optical confinement and device miniaturization compared with high-index semiconductor platforms \cite{Pelucchi2022Thepotential}. Consequently, low-loss routing, active control, nonlinear frequency conversion, and quantum-state generation often need to be distributed across multiple materials and heterogeneous integration schemes.\\
	This fragmentation is particularly restrictive for photonic quantum computing, where single photons and entangled photon pairs constitute important resources for quantum interference, teleportation, entangling operations, and the generation of multiphoton states \cite{Wang2020Integrated,Caspani2017Integrated}. Spontaneous parametric down-conversion (SPDC) is among the most established and versatile processes for generating correlated and entangled photon pairs. Its efficiency depends critically on the second-order nonlinear susceptibility and the strength of the light-matter interaction. Integrating SPDC into high-index nanophotonic structures can therefore increase source brightness, reduce the required interaction length and pump power, and enable compact sources within densely integrated quantum-photonic structures \cite{Shi2024Efficient,Weissflog2024Nonlinear,Fan2025Enhanced}. A low-loss material combining strong optical confinement with a large intrinsic second-order nonlinear response would therefore provide an important foundation for scalable photonic quantum technologies.\\
	Gallium phosphide (GaP) has emerged as a prime candidate in this context \cite{Wang2024Review}. As a high-index, non-centrosymmetric III-V semiconductor, it combines strong optical confinement with pronounced second- and third-order nonlinear responses. Its high refractive index ($n>3$) enables subwavelength confinement and small optical mode volumes \cite{Cambiasso2017Bridging}, while its wide indirect bandgap ($E_g=\qty{2.26}{\electronvolt}$) supports low linear absorption across broad spectral ranges from the visible to the infrared \cite{Wilson2020Integrated,Kuznetsov2025Ultrabroadband}. These properties have enabled efficient nonlinear frequency conversion \cite{Anthur2021Secondharmonic,Anthur2020Continuous,Yang2024Secondharmonic}, Kerr nonlinear processes \cite{Xie2019Nonlinear}, and electro-optic and piezoelectric functionalities \cite{Stockill2022Ultralownoise}. This combination positions GaP as a highly versatile material for compact nonlinear, nanophotonic, and quantum-optical devices.\\
	The broader exploitation of these capabilities has, however, been limited by the lack of a scalable GaP-on-insulator platform that combines single-crystalline material quality with transparent, low-index host substrates. The large refractive-index contrast between the GaP device layer and the underlying insulator enables efficient waveguiding, strong optical confinement, and high-quality-factor resonators needed for integrated nanophotonic devices. Existing fabrication strategies seek to achieve this platform while balancing optical performance, process complexity, and scalability. Direct top-down patterning of bulk GaP substrates yields high-quality nanostructures for free-space nanophotonics but lacks the refractive-index contrast required for integrated waveguide devices \cite{Cambiasso2017Bridging,Anthur2020Continuous,Yang2024Secondharmonic}. Alternatively, sputter-deposited amorphous GaP films offer excellent substrate flexibility and large-area fabrication, yet exhibit higher optical losses and inferior nonlinear optical performance than single-crystalline GaP \cite{Tilmann2020Nanostructured}. Single-crystalline GOI has primarily been realized either by epitaxial growth on sacrificial layers followed by membrane release \cite{Rivoire2009Secondharmonic,Lake2016Efficient} or by wafer bonding of epitaxially grown GaP layers onto low-index substrates with subsequent removal of the growth substrate and sacrificial layers \cite{Wilson2020Integrated,Kuznetsov2025Ultrabroadband}. While membrane platforms are well suited for suspended micro- and nanocavities, wafer-bonded GOI has become the state-of-the-art platform for integrated nonlinear photonics, albeit at the expense of sophisticated epitaxial layer structures and complex bonding and substrate-removal processes limiting scalability.\\
	Ion-slicing offers a scalable route to high-quality single-crystalline GOI substrates \cite{Bruel1995Silicon,Maleville1997Waferbonding,Moriceau2012SmartCut}. In this approach, light-ion implantation defines a buried cleavage plane, while subsequent wafer bonding and thermal annealing induce controlled exfoliation of a thin crystalline film onto a low-index host substrate. Originally developed for silicon-on-insulator technology \cite{Bruel1995Silicon}, ion-slicing has since been extended to functional crystals such as lithium niobate \cite{Rabiei2004Optical}, lithium tantalate \cite{Tauzin2008LiTaO3, Liu2008Fabrication}, silicon carbide \cite{DiCioccio1996Silicon}, and gallium nitride \cite{Huang2017Investigation}. The method provides precise, implantation-defined thickness control and enables the integration of single-crystalline layers onto a wide range of host substrates \cite{Tong1999Waferbonding,Moriceau2012SmartCut}. Ion-assisted transfer of GaP onto borosilicate glass was previously reported \cite{Ganjoo2020Transfer}, however, the crystalline quality and optical properties of the transferred layers, as well as post-transfer annealing and surface-polishing, were not demonstrated. A comprehensive demonstration of ion-sliced, single-crystalline GOI substrates suitable for nanophotonic and nonlinear-optical integration therefore remains outstanding.\\
	Here, we apply ion-slicing to fabricate single-crystalline GOI substrates using energetic He$^{+}$ implantation and wafer bonding, as illustrated in Figure~\ref{fig:ionslicing}(a). Beyond demonstrating controlled GaP layer transfer, we investigate the structural recovery, surface quality, as well as linear and nonlinear optical properties of the resulting films. We further investigate two complementary wafer bonding approaches: anodic bonding and plasma-enhanced direct bonding. Anodic bonding is more tolerant to surface imperfections but relies on alkali-containing glass substrates, limiting its compatibility with CMOS process flows \cite{Knowles2006anodic}. In contrast, plasma-enhanced direct bonding is compatible with CMOS processing and a broad range of substrate materials, albeit at the expense of more stringent surface-quality requirements \cite{Suni2002Effects,SchojbergHenriksen2002Oxide}.
	\section*{Results and Discussion}\label{sec:results}
	\subsection*{Ion Irradiation and Wafer Bonding}\label{subsec:ionslicing}
\begin{figure*}[ht]
		\centering
		\includegraphics[width=0.9\textwidth]{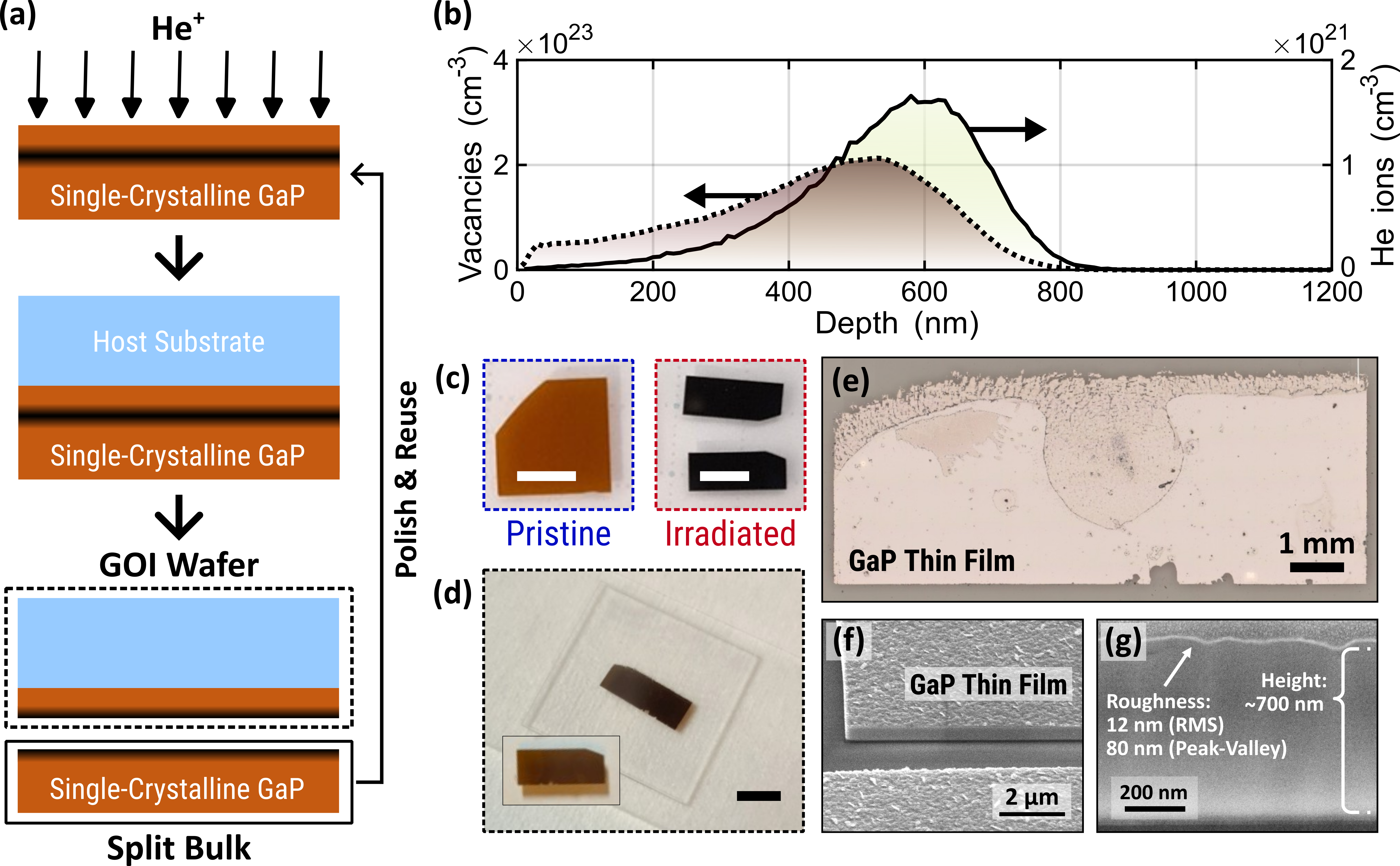}
		\caption{(a) Schematic for the ion-slicing of GaP using He$^{+}$-ions, showing ion irradiation (top), wafer bonding (middle), and splitting (bottom). (b) Depth-dependent static vacancy (left axis) and He-ion (right axis) concentration profile for GaP irradiated by \qty{100}{\kilo\electronvolt} He$^{+}$-ions with a fluence of \qty{5e16}{\per\centi\metre\squared}. (c) Photograph of typical GaP samples before (left) and after (right) ion irradiation. (d) Photograph of a typical as-bonded GOI substrate fabricated by anodic bonding. The inset shows the split GaP sample. The scale bars in (c,d) indicate \qty{5}{\milli\metre}. (e) Optical micrograph of the as-bonded GOI substrate. (f) Perspective and (g) cross-section view scanning electron micrographs of the same sample.}
		\label{fig:ionslicing}
	\end{figure*}

	We irradiated (100)-, (110)-, and (111)-oriented GaP single crystals with \qty{100}{\kilo\electronvolt} He$^+$ ions using an ion fluence of \qty{5e16}{\per\centi\metre\squared}. The irradiation was performed at room temperature and under an incidence angle of \qty{7}{\degree} to avoid channeling effects. The static ion implantation profile can be approximated by Monte Carlo simulations, and Figure~\ref{fig:ionslicing}(b) shows the respective projected, depth-dependent vacancy and He-ion concentrations calculated using the simulation package SRIM -- Stopping and Range of Ions in Matter \cite{Ziegler2010SRIM}.\\
	The irradiation profile is tailored to promote exfoliation of the GaP layer at a well-defined depth. In other crystalline materials, particularly silicon, helium-induced ion-slicing is understood to result from the coupled evolution of implantation damage and helium accumulation during post-implantation annealing.
	Implantation-induced defects provide preferential trapping and nucleation sites for helium, while the subsequent growth and coalescence of helium-filled bubbles or platelets generate highly localized stresses that initiate microcracks and ultimately lead to layer exfoliation \cite{Maleville1997Waferbonding, Moriceau2012SmartCut, Roth2006Compositional}. Consequently, splitting occurs typically where implantation damage and helium concentration overlap and cavity formation is most pronounced, rather than being determined by either quantity alone \cite{Aspar2001Thegeneric, Roth2006Compositional}. Although the microscopic mechanisms governing helium-induced exfoliation have not yet been investigated in comparable detail for GaP, SRIM simulations provide a physically motivated estimate of the implanted region, yielding an anticipated exfoliation depth near \qty{600}{\nano\metre}.\\
	Figure~\ref{fig:ionslicing}(c) shows photographs of (100)-oriented GaP single crystals before (left) and after (right) He$^{+}$-irradiation. Following irradiation, the crystals become visibly opaque, changing from bright orange to black owing to the formation of defect-related absorption and scattering centers associated with implantation-induced atomic disorder and He-filled nanobubbles. The as-implanted samples were subsequently wafer-bonded as described in the Experimental Section. Figure~\ref{fig:ionslicing}(d) shows a (100) GOI substrate immediately after anodic wafer bonding together with the split bulk GaP crystal (inset). Notably, crystal splitting occurs during the wafer bonding process that includes heating to \qty{350}{\degreeCelsius}. The optical microscope image of the same GOI substrate shown in Figure~\ref{fig:ionslicing}(e) demonstrates successful material transfer by exfoliation over nearly the entire sample area, with large regions remaining intact as continuous thin films. Scanning electron micrographs taken at delaminated areas in Figure~\ref{fig:ionslicing}(e,f) further confirm the transfer of a GaP thin film onto the host substrate. The cross-sectional view in Figure~\ref{fig:ionslicing}(f) reveals a film thickness of approximately \qty{700}{\nano\metre}, while the thickness across different samples typically ranges from \qtyrange{600}{700}{\nano\metre}. These values exceed the splitting depth predicted by SRIM simulations, suggesting that the actual implantation profile may extend deeper than estimated by the amorphous-target model, for example owing to residual ion channeling or uncertainties in the stopping-power description of GaP. Direct depth profiling would be required to clarify the relationship between the He distribution, implantation-induced damage, and cleavage-plane formation, which lies beyond the scope of the present work. GOI substrates were further fabricated from (110)- and (111)-oriented bulk GaP crystals using both anodic and plasma-enhanced direct bonding, as demonstrated in Section~S1 of the Supporting Information.
	\subsection*{Structural Analysis}
	Rutherford backscattering spectrometry in channeling geometry (RBS/C) was performed with \qty{2.3}{\MeV} He$^{+}$ ions to assess the structural quality of the as-bonded GOI substrates (see Experimental Section for details). Figure~\ref{fig:StructuralCharacterization}(a) shows the resulting RBS spectra of an as-bonded GOI substrate fabricated from a (100)-oriented bulk GaP crystal by anodic bonding. In the random orientation (black circles), two dominant elemental edges are observed at approximately \qty{1.83}{\MeV} and \qty{1.37}{\MeV}, corresponding to the surface Ga and P signals, respectively. The vertical dashed lines mark their expected onset energies and agree well with the experimental data, while the finite widths of the Ga and P signals reflect the thin-film geometry of the transferred GaP layer. Additional signals from the underlying borosilicate glass substrate, including silicon (Si), aluminum (Al), sodium (Na), and oxygen (O), appear below approximately \qty{1}{\MeV}. As these elements are buried beneath the GaP film, their edge energies are shifted to lower values by the energy loss of the incident and backscattered He$^{+}$ ions while traversing the overlying GaP layer (see vertical dashed lines). Comparison with simulated RBS spectra yields a GaP film thickness of approximately \qty{595}{\nm} and a Ga:P stoichiometric ratio of 1:1 within the range of error.\\
	%
	%
	When the sample is aligned with a major channeling direction (blue circles), the Ga and P signals from the transferred layer exhibit a pronounced reduction in backscattering yield, demonstrating that the crystalline order is largely preserved on a macroscopic scale (beam diameter $\sim$\qty{1}{\milli\meter}) following ion irradiation and transfer onto the borosilicate glass substrate. The channeling contrast nevertheless varies with depth: it is well defined in the deeper region of the GaP layer but markedly reduced near the surface. This near-surface region coincides with the implantation-damaged layer generated during ion slicing (compare with Figure~\ref{fig:ionslicing}(b)), indicating an elevated defect concentration while retaining substantial crystalline order. As the damage is confined to approximately the upper \qty{150}{\nm} of the film, it can be removed by subsequent etching and polishing to expose the underlying high-quality single-crystalline GaP.
	\subsection*{Annealing and Polishing}
\begin{figure*}[ht]
		\centering
		\includegraphics[width=0.95\textwidth]{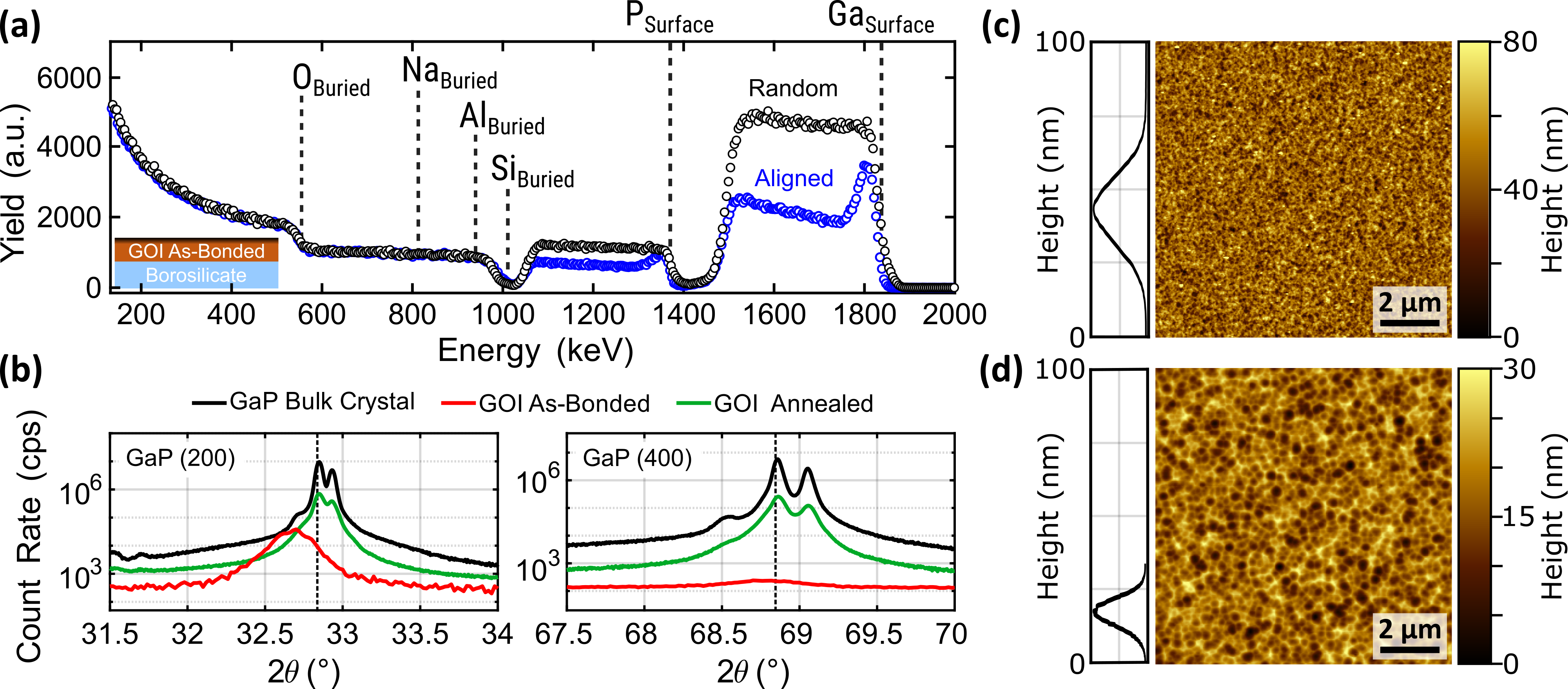}
		\caption{Measured RBS spectra in random (black circles) and channeling (blue circles) geometry for a GOI substrate fabricated by anodic bonding. (b) XRD spectra of the (200) (left) and (400) (right) Bragg peaks of GaP measured for bulk GaP (black curves) and GOI substrates before (red curves) and after annealing (green curves). Atomic force microscopy measurements of a GOI substrate (c) before and (d) after ion beam milling. The graphs on the left show the corresponding height histograms.}
		\label{fig:StructuralCharacterization}
	\end{figure*}

	Thermal annealing followed by surface polishing was performed to improve the crystalline and surface quality of the fabricated GOI substrates. The samples were annealed at \qty{500}{\degreeCelsius} for \qty{210}{\minute} in a vacuum furnace at a pressure of approximately \qty{e-6}{\milli\bar}. The influence of thermal annealing on the crystalline quality of the GOI substrates was investigated by X-ray diffraction (XRD). Figure~\ref{fig:StructuralCharacterization}(b) shows the corresponding symmetric $\theta/2\theta$ scans of a (100)-oriented bulk GaP crystal (black curves) and GOI substrates before (red curves) and after annealing (green curves). The left and right panels display the angular ranges around the expected (200) and (400) Bragg diffraction peaks, respectively, with the corresponding Bragg angles marked by vertical dashed lines.
	The bulk GaP crystal exhibits sharp diffraction peaks in excellent agreement with the expected Bragg angles. In contrast, the as-bonded GOI substrate shows a strongly broadened and shifted (200) peak, while the higher-order (400) peak is nearly undetectable. This indicates a substantial degradation of the long-range crystalline order following ion implantation, accompanied by significant lattice strain and defect accumulation.\\
	Following thermal annealing, the Bragg peaks are largely restored and no significant peak shifts remain, indicating recovery of the average lattice spacing. The diffraction peaks nevertheless remain moderately broader than those of the pristine bulk crystal, indicating residual inhomogeneous lattice strain and crystalline defects. The count rates measured for the GOI substrates before and after annealing are one to two orders of magnitude lower than those of the bulk reference, primarily because the transferred GaP layers are thinner than the X-ray penetration depth in GaP.\\ \\
	The annealed GOI substrates were subsequently polished by Ar$^{+}$ ion milling, as detailed in the Experimental Section. Figures~\ref{fig:StructuralCharacterization}(c,d) compare the surface morphology of the fabricated GOI substrates before and after polishing, respectively, as measured by atomic force microscopy (AFM). In the as-bonded GaP films, ion-range straggling (Figure~\ref{fig:ionslicing}(b)) broadens the depth region over which He-bubble formation and microcrack propagation occur, resulting in a rough surface after exfoliation. The corresponding AFM measurement yields a peak-to-valley roughness of approximately \qty{80}{\nano\metre} and a root-mean-square roughness of \qty{12}{\nano\metre}. Uniform ion milling with \qty{400}{\electronvolt} Ar$^{+}$ ions at a shallow incidence angle removes material from the GaP surface while simultaneously smoothing the film. After \qty{20}{\minute} of milling (Figure~\ref{fig:StructuralCharacterization}(d)), the surface morphology is visibly improved (see scales in Figure~\ref{fig:StructuralCharacterization}(c,d)), with the peak-to-valley and root-mean-square roughness reduced to approximately \qty{30}{\nano\metre} and \qty{4}{\nano\metre}, respectively.
	\subsection*{Optical Properties}
\begin{figure*}[ht]
		\centering
		\includegraphics[width=0.98\textwidth]{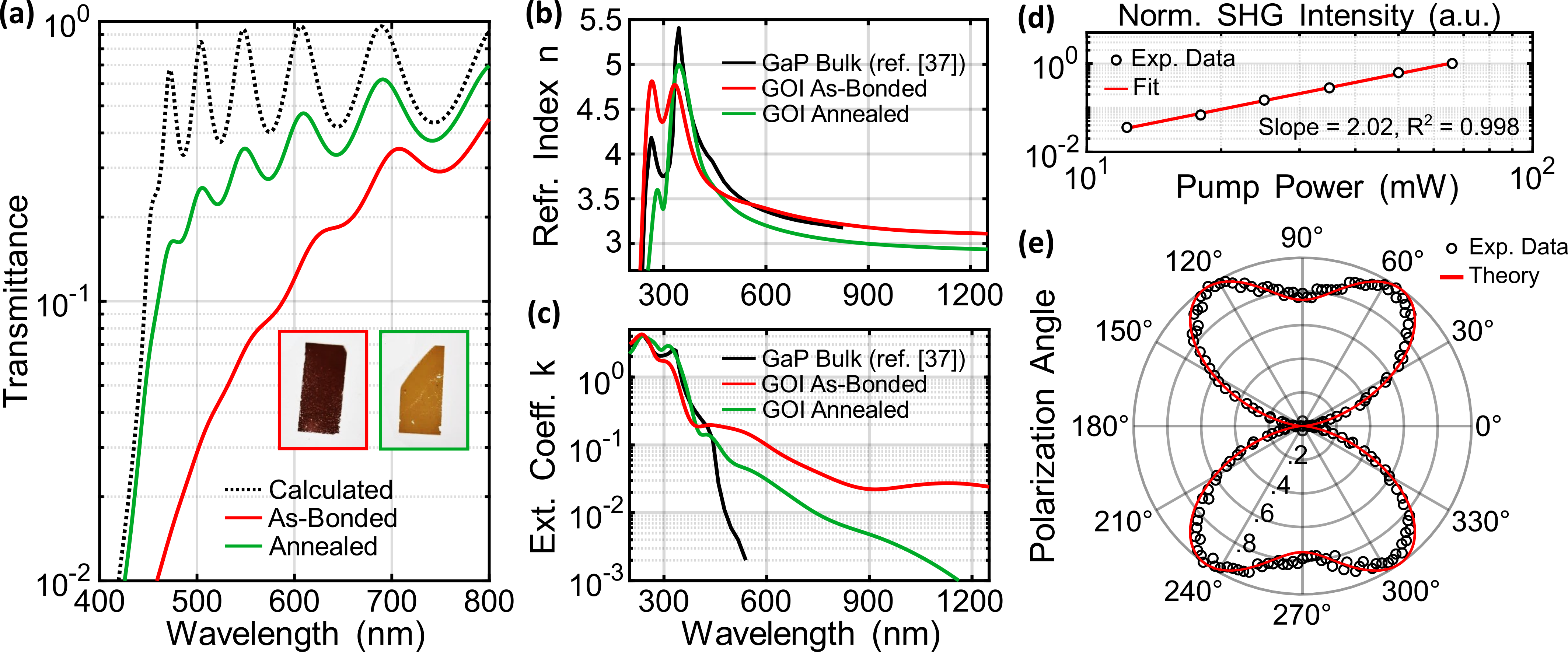}
		\caption{(a) White light transmittance spectra measured for as-bonded (red curve) and annealed (green curve) GOI substrates, as well as the calculated transmittance of a \qty{634}{\nm} thick GaP thin film on borosilicate glass. (b) Refractive index and (c) extinction coefficient spectra obtained by multi-angle spectroscopic ellipsometry from an as-bonded (red curves) and an annealed (green curves) GOI substrate as well as literature data of bulk GaP taken from \cite{Aspnes1983Dielectric}. (d) Excitation power dependence and (e) linear polarization angle dependence of the normalized SHG intensity from a GOI substrate fabricated from a (110)-oriented GaP bulk crystal.}
		\label{fig:OpticalProperties}
	\end{figure*}

	%
	%
	Figure~\ref{fig:OpticalProperties}(a) shows the measured linear transmittance spectra of an as-bonded (red curve) and an annealed (green curve) GOI substrate before polishing. Both spectra exhibit the characteristic Fabry-Pérot oscillations of a layered thin-film structure. The as-bonded sample shows a comparatively low overall transmittance, consistent with its dark brown appearance in the photograph (inset, red frame), indicating a broadband absorption across the measured spectral range. Accordingly, its transmittance remains below 10\% for wavelengths shorter than approximately \qty{600}{\nm}.
	The transmittance increases substantially following thermal annealing, as also reflected by the light orange appearance of the sample (inset, green frame). In addition, a pronounced absorption edge emerges, resembling that of bulk GaP, with transmittance values remaining above 10\% for to wavelengths longer than approximately \qty{450}{\nm}. For comparison, the black dashed curve shows the calculated transmittance of a \qty{634}{\nm} thick GaP thin film, obtained with the transfer matrix method \cite{Yeh1977Electromagnetic} and using the optical constants of bulk GaP taken from \cite{Aspnes1983Dielectric}. The measured spectrum of the annealed sample closely follows the calculated response, indicating that the optical properties approach those of bulk single-crystalline GaP. The remaining reduction in transmittance can be attributed to residual optical absorption and scattering losses associated with stable implantation-induced defects and the unpolished surface.\\
	Spectroscopic ellipsometry was further employed to quantify the effect of thermal annealing on the optical constants of the transferred GaP thin films. Figures~\ref{fig:OpticalProperties}(b,c) show the measured refractive index $n$ and extinction coefficient $k$, respectively, of a GOI substrate before (red curves) and after annealing (green curves) alongside the corresponding results for bulk GaP (black curves) taken from \cite{Aspnes1983Dielectric}. The measured samples exhibit a high refractive index of approximately $n=3$ or higher across the measured spectral range. Following annealing, the extinction coefficient of the GOI substrate decreases markedly at wavelengths above approximately \qty{400}{\nm}, indicating a substantial reduction in implantation-induced optical losses.\\
	Owing to the indirect band gap of bulk GaP, absorption near the fundamental transition at approximately \qty{550}{\nm} remains comparatively weak, whereas the pronounced increase in absorption below approximately \qty{400}{\nm} is associated with higher-energy direct interband transitions. Similar to bulk GaP, the annealed GaP film therefore exhibits a broad transparency window across much of the visible spectrum. This optical recovery is consistent with the change in the visual appearance of the sample and the increased transmittance shown in Figure~\ref{fig:OpticalProperties}(a), while the narrowing of the XRD peaks in Figure~\ref{fig:StructuralCharacterization}(b) confirms a simultaneous recovery in crystalline order.\\ \\
	%
	A key optical property of single-crystalline GaP is its strong second-order nonlinear response, which arises from the absence of inversion symmetry in its zinc-blende crystal structure. This enables efficient frequency conversion processes such as SPDC and second-harmonic generation (SHG) with the conversion efficiency of both processes governed by the relevant components of the second-order nonlinear susceptibility tensor \cite{Weissflog2024Nonlinear,Anthur2021Secondharmonic}. Owing to the crystal symmetry, however, the bulk electric-dipole contribution to SHG vanishes in the exact forward direction under normal-incidence excitation of a (100)-oriented crystal, whereas suitable polarization configurations in (110)- and (111)-oriented crystals permit forward-emitted SHG. Importantly, ion slicing provides access to different GaP crystal orientations without requiring orientation-matched growth substrates, as the desired orientation is defined by the bulk donor crystal. We therefore fabricated GOI substrates from (110)- and (111)-oriented GaP and characterized their second-order nonlinear response in transmission, as detailed in the Experimental Section. Figure~\ref{fig:OpticalProperties}(d) shows the detected signal of the (110) GOI substrate at twice the pump frequency as a function of incident pump power, exhibiting a clear quadratic power dependence that is characteristic of SHG.\\
	Figure~\ref{fig:OpticalProperties}(e) shows the normalized SH intensity from the (110) GOI substrate as a function of the linear polarization angle of the pump beam (see Section~S2 in the Supporting Information for the corresponding measurements of the (111) GOI substrate). The measured angular dependence (black circles) closely follows the theoretical response (red curve), obtained from the non-zero components of the second-order susceptibility tensor of a (110)-oriented zinc-blende crystal \cite{Anthur2021Secondharmonic}. The close agreement between experiment and theory further indicates that the crystallographic orientation and macroscopic crystal symmetry are well preserved in the GaP thin film after layer transfer. In particular, the absence of pronounced distortions or symmetry breaking in the polarization pattern suggests that implantation-induced disorder and strain-related lattice distortions are substantially reduced upon annealing, resulting in a nonlinear optical response of the GOI substrate approaching that of pristine single-crystalline GaP. Furthermore, following annealing the GOI substrates yield a substantially increased SH signal (see Section~S2 in the Supporting Information) owing to the reduced optical absorption at the pump and the SH wavelength.
	\section*{Conclusion}\label{sec3}
	We have demonstrated the fabrication of single-crystalline GaP thin films on transparent, low-index substrates by ion slicing with energetic He$^{+}$ ions in combination with wafer bonding. By tailoring the implantation energy and fluence, the ion penetration depth and exfoliation behavior of irradiated bulk GaP were controlled, enabling the transfer of thin GaP layers onto amorphous host substrates by anodic bonding and plasma-enhanced direct wafer bonding. The compatibility of the process with different substrate materials, crystal orientations, and bonding conditions was demonstrated and highlights its potential for integration into scalable and CMOS-compatible fabrication workflows.\\
	Channeling RBS and XRD measurements confirmed that the transferred GaP layers retain their single-crystalline structure. Implantation-induced disorder and the associated optical absorption were substantially reduced by annealing at \qty{500}{\degreeCelsius} and subsequent surface polishing. The resulting GOI substrates exhibited high linear optical quality, while polarization-resolved SHG measurements on a (110)-oriented GaP layer revealed a second-order nonlinear response approaching that of pristine bulk GaP.\\
	This fabrication platform enables high-quality single-crystalline GaP thin films to be integrated with transparent, low-index substrates, thereby providing the optical confinement and material quality required for waveguide-based integrated photonics as well as free-space nanoantennas and metasurfaces. More broadly, the orientation flexibility of ion slicing offers access to crystal cuts optimized for specific linear and nonlinear and quantum optical functionalities.
	\section*{Experimental Section}\label{sec4}
	\subsection*{Ion-Slicing}
	Bulk GaP crystals were subjected to He\textsuperscript{+} ion implantation at room temperature using a fluence of $5 \times 10^{16}\,\text{cm}^{-2}$, an acceleration energy of \qty{100}{\keV}, and a tilt angle of \ang{7} with respect to the surface normal. Prior to homogeneous irradiation, the GaP crystals and the host substrates were cleaned in an ultrasonic acetone bath for \qty{5}{\minute}, rinsed with isopropanol, blow-dried with nitrogen, and subsequently cleaned by a \qty{100}{\watt} O$_2$ plasma (Diener Electronic Zepto). The surfaces of the cleaned GaP crystal and the host substrate were then brought into intimate contact under applied pressure to maximize the bonding area, while the process chamber was evacuated to minimize interfacial voids and trapped air pockets. Bonding was performed using a SUSS MicroTec SB6 wafer bonder.\\
	For anodic wafer bonding, the initial plasma cleaning duration was \qty{5}{\minute} and for the subsequent bonding process a temperature of \qty{350}{\degreeCelsius} and an applied voltage of \qty{600}{\volt} were used. For plasma-activated direct bonding, the initial plasma cleaning duration was \qty{15}{\minute} and the treated substrates were then kept in deionized water. The subsequent bonding was carried out at a temperature of \qty{400}{\degreeCelsius} for \qty{45}{\minute}. The transferred films were then annealed at \qty{500}{\degreeCelsius} for \qty{3.5}{\hour} using a temperature ramp of \qty{3}{\kelvin/\minute} for heating and cooling. The subsequent polishing was carried out using a homogeneous Ar\textsuperscript{+} ion beam incident at an angle of \ang{15} relative to the surface plane, with an ion energy of \qty{400}{\eV} (Oxford Instruments, Ionfab 300).
	\subsection*{Thin Film Characterization}
	Surface and cross-sectional morphology were examined by scanning electron microscopy using an FEI Helios 600i NanoLab dual-beam system equipped with a focused ion beam. Surface roughness and film thickness were characterized by atomic force microscopy. Crystallinity and phase composition were assessed by $\theta/2\theta$ X-ray diffraction measurements (Bruker, D8 Advance) using Cu-K$_{\alpha}$ radiation ($\lambda = \qty{1.5418}{\angstrom}$) as well as by Rutherford backscattering spectrometry, employing a \qty{2.3}{\mega\electronvolt} He$^{+}$ ion beam at a scattering angle of \qty{170}{\degree}.\\
	Spectroscopic ellipsometry was performed using a multi-angle ellipsometer (J.A. Woollam, M-2000D). Optical transmission spectra were recorded using a spectrophotometer (Cary 5000). The nonlinear optical measurements were performed using a custom-built nonlinear microscopy setup. The excitation source was a Chameleon Compact optical parametric oscillator (OPO, Coherent Corp.) pumped by a Chameleon Ultra II laser (Coherent Corp.). The OPO output at a wavelength of 1200 nm was focused onto the sample using a Mitutoyo objective with a numerical aperture of 0.2. The excitation power and polarization were controlled using two half-wave plates and a linear polarizer positioned before the focusing objective. The generated second-harmonic signal was collected and projected onto a camera using a lens system. A short-pass filter placed between the two lenses suppressed the residual fundamental radiation before detection.
	%
	%
	%
	%
	\subsection*{Acknowledgments}
	We thank Parisa Mirzaei and Andreas Bätz for assistance with ion irradiation and analysis procedures, Oliver Rüger for assistance with preparing FIB-SEM cross-section images, Maximilian Wei\ss{}flog for assistence with optical profilometry, and Daniel Voigt for performing the ion milling procedures. This work was funded by the German Federal Ministry of Research, Technology and Space (BMFTR) within the project GOI-4-IQ-NANO (Research Program Quantum Systems, Contract Number~13N17107), the European Research Council through the ERC Synergy Grant "ATHENS" (Grant Agreement No.~101167540), the International Research Training Group (IRTG) 2675 "Meta-Active - Tailored metasurfaces - generating, programming and detecting light" (Project Number~437527638), and the RTG~3014 "PhInt - Photo-Polarizable Interfaces and Membranes" funded by the Deutsche Forschungsgemeinschaft (DFG, Project Number~521747072).
	\subsection*{Conflicts of Interest}
	The authors declare no conflicts of interest.
	\subsection*{Data Availability Statement}
	The data that support the findings of this study are available from the corresponding authors upon reasonable request.
	\bibliographystyle{unsrt}
	\bibliography{manuscript_references}
	\subsection*{Supporting Information}
	Additional supporting information can be found in the Supporting Information
	section.
	\newpage
	\includepdf[pages={1-3}]{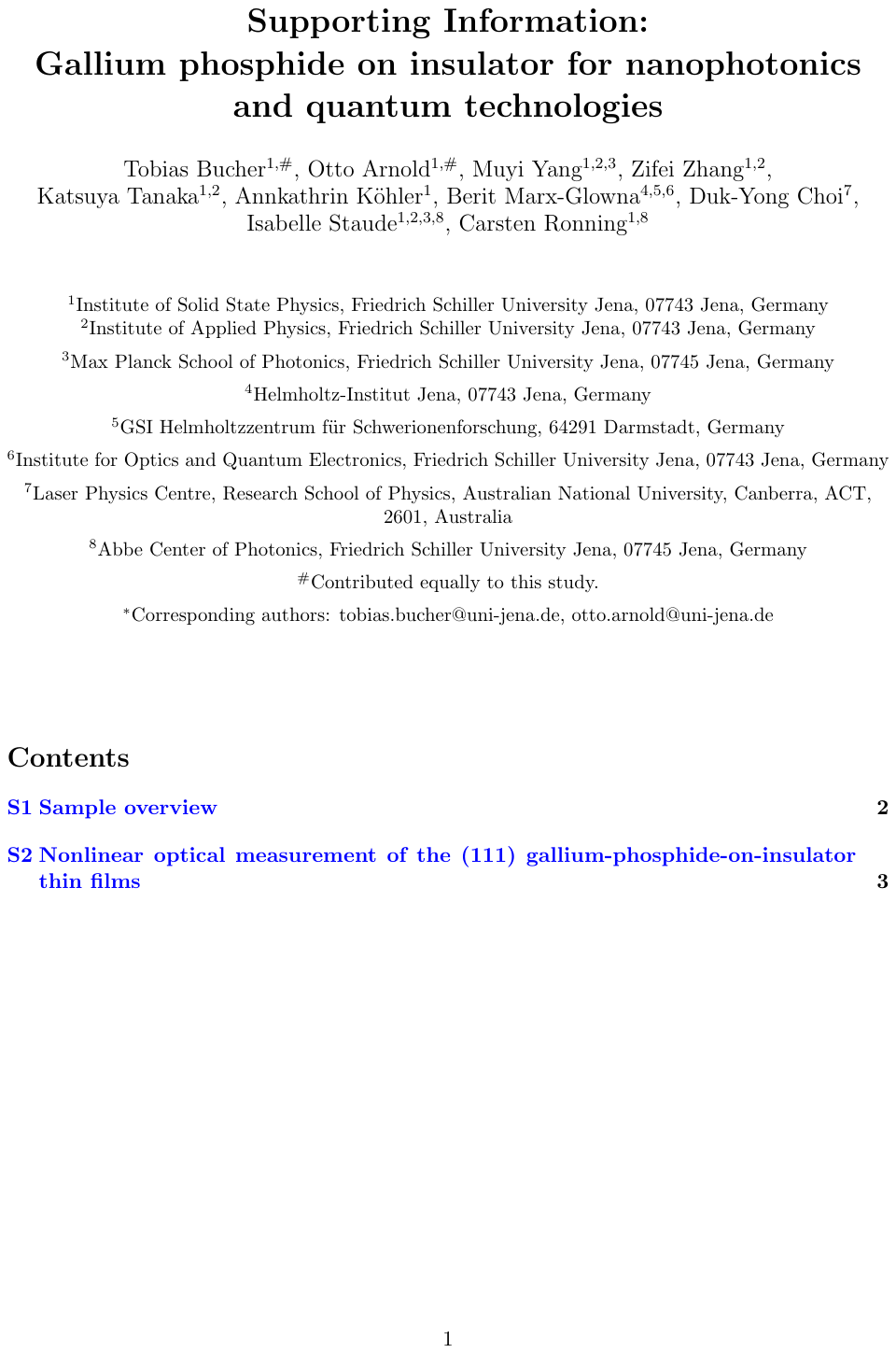}
	
\end{document}